\documentclass[letterpaper, 10pt, conference]{ieeeconf}
\IEEEoverridecommandlockouts
\usepackage{amsfonts, amssymb, amsmath} 
\usepackage{graphicx}
\usepackage{epstopdf}
\usepackage{tikz}
\usetikzlibrary{decorations.pathreplacing}
\usetikzlibrary{arrows,positioning,automata,shadows,fit,shapes}

\begin{document}

\title{Virtual Modules in Discrete-Event Systems: \\ Achieving Modular
Diagnosability} \author{Dmitry Myadzelets$^{*,1,2}$, Andrea Paoli$^{1,3}$ 
% --- Draft --- \today
\thanks{$^{1}$Center for Research on Complex Automated Systems
(CASY), DEI, University of Bologna, Viale Pepoli 3/2, 40123, Bologna, Italy}
	\thanks{$^{2}$E-mail: {dmitry.myadzelets@gmail.com}}
	\thanks{$^{3}$E-mail: {andrea.paoli@unibo.it}}
} \maketitle

% To meet requirements of EU funding
% http://eacea.ec.europa.eu/about/eacea_logos_en.php "This project has been
% funded with support from the European Commission. This publication reflects
% the views only of the author, and the Commission cannot be held responsible
% for any use which may be made of the information contained therein."
\begin{figure}[!b]
\begin{tabular}{l p{60mm}}
 	\includegraphics[height=10mm]{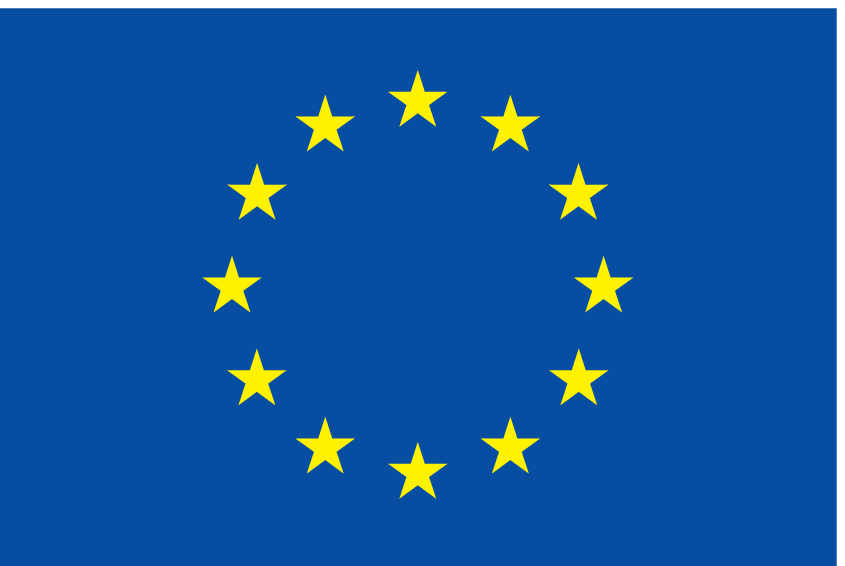}
 	& \vspace{-10mm} \footnotesize
 	$^{*}$With the support of the Erasmus Mundus Action 2 programme of the
 	European Union
\end{tabular}
\end{figure}

\begin{abstract}
This paper deals with the problem of enforcing modular diagnosability for
discrete-event systems that don't satisfy this property by their natural
modularity. We introduce an approach to achieve this property combining existing
modules into new virtual modules. An underlining mathematical problem is to find
a partition of a set, such that the partition satisfies the required property.
The time complexity of such problem is very high.
To overcome it, the paper introduces a structural analysis of the system's
modules. In the analysis we focus on the case when the modules participate in
diagnosis with their observations, rather then the case when indistinguishable
observations are blocked due to concurrency.
\end{abstract}

\begin{keywords}
Discrete Event Systems, Modular Structure, Distributed Diagnosability
\end{keywords}

\newtheorem{assumption}{Assumption}
\newtheorem{definition}{Definition}
\newtheorem{conjecture}{Conjecture}
\newtheorem{lemma}{Lemma}
\newtheorem{corollary}{Corollary}
\newtheorem{example}{Example}
\newtheorem{theorem}{Theorem}

% %%%%%%%%%%%%%%%%%%%%%%%%%%%%%%%%%%%%%%%%%%%%%%%%%%%%%%%%%%%%%%%%%%%%%%%%%%%%%%
\section{Introduction}
% Raffly: What we are talking about and what problem exist there in general. How
% the problems are solved. What we suggest exactly and what is that for.
% How this paper is structured.

Discrete-Event Systems (DES) has successfully concurred significant area in the
systems engineering discipline due to their enormous capabilities of designing
and managing complex systems. While ``real-world'' systems are growing in scale,
the solutions provided by DES have to evolve to tackle the increasing complexity
issues. For that reason the development of solutions relying on the fact that
the most of complex systems have naturally modular structure has been under
focus for the last two decades. Particularly, the task of design verification
and diagnosis with respect to undesired behaviour, commonly refereed as to
faulty behaviour of discrete systems, has a fairly developed theory nowadays.
In this paper we consider the automata framework for diagnosability analysis,
where a behaviour of DES is modeled by regular languages and represented by
automata.

Diagnosability analysis requires to verify if one can detect if the system
executes a faulty behaviour, i.e. a fault occurred, and to verify if one
can isolate a certain type of fault from other faults. This analysis implies
that the system's behaviour can be observed only partially. In DES built from
more then one modules it may be necessary to verify if the faults originated
from one module can be detected by observing only the same module, or by
observing only other modules, or under other implications with respect to the
possible flow of observations. Moreover, the verification of a modular system is
preferred to be done without composing its entire model from the system's
components since such composition may be not even feasible to perform due to 
correspondent high computational burden.

The approaches aimed to solve the problem of diagnosability verification
consider different architectures of DES and differ with respect to some
implications they take into account, i.e. either they require an entire model of
the system or not, what information is presented to each observation spot if any,
and etc. In this paper we use a classification as in \cite{su_global_2005} where
a diagnosability approach can be \emph{centralized},
\emph{decentralized} or \emph{distributed}. The centralized approach is
presented in \cite{sampath_diagnosability_1995} and, with improved complexity,
in \cite{jiang_polynomial_2001} and \cite{yoo_polynomial-time_2002}.
The decentralized approach can be found in \cite{debouk_coordinated_1998},
\cite{pencole_formal_2005}, \cite{qiu_decentralized_2006},
\cite{wang_diagnosis_2007} and others.
The distributed approach is presented in \cite{su_distributed_2002}, and a
related notion of modular diagnosability is introduced in
\cite{contant_diagnosability_2006}. We briefly review all the approaches in the
next section.

The original contribution of this paper can be summarized as follows. We
consider the distributed approach for a DES with modular structure, i.e. no
entire model of the system is presented and the fault diagnosis procedure
assumes that observation spots can not communicate. As the starting point we
consider the definition of modular diagnosability property and the correspondent
verification algorithm from \cite{contant_diagnosability_2006}. We assume that
the systems' modules are given by its designer. The design may reflect a
physical or functional structure of the system, or may follow other underlining
design principals, which make the modules \emph{natural} for the designer.
Thus, we assume that preserving the systems' modularity as close to the initial
structure as possible is required. We investigate the case when the system is
not modular diagnosable initially, but the modules of the system can be composed
into new \emph{virtual} modules in order to force the modular diagnosability
property. We study how to choose the system's modules for the composition. For
this goal a structural analysis of the system is introduced. 

We refer to \cite{ye_optimized_2010} as to one of the recent works addressing
diagnosability problem for distributed approach. The work introduces a notion of
\emph{regional diagnosability} exploiting a subset of the modules. The authors
do not address modular diagnosability property, however, we may consider this
work as a similar due to the correlated notion, and optimisation techniques,
which can be applied in our approach.

This paper is organized as follows. Section \ref{sec:Preliminaries} covers the
necessary notation and describes the diagnosability problem. Section
\ref{sec:Diagnosability} reviews diagnosability verification of a modular
system. In Section \ref{sec:Proposal} we focus on diagnosability by virtual
modules, and analysis of system's module structure with respect to the faulty
behaviour. The Section \ref{sec:Example} shows an example. The last Section
\ref{sec:Conclusion} concludes the current results and discusses possible
directions for the further research.

%%%%%%%%%%%%%%%%%%%%%%%%%%%%%%%%%%%%%%%%%%%%%%%%%%%%%%%%%%%%%%%%%%%%%%%%%%%%%%%
% Move this picture to a place it looks better, if necessary.  
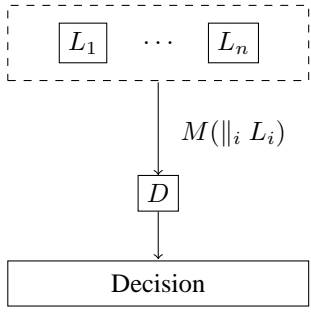
\begin{figure}[t]
\centering
\begin{tikzpicture}
	\node[draw]		(0)	at (1, 3) 		{$L_1$};
	\node[]			(1)	[right of = 0]	{$\ldots$};
	\node[draw]		(2)	[right of = 1]	{$L_n$};
	\draw[dashed] (0, 2.5) rectangle (4, 3.5);
	\node[]			(3) at (2, 2.6) {};
	\node[draw]		(4) at (2, 1) {$D$};
	\draw[->] (3) -- (4);
	\node[]	at (3, 1.8) {$M(\parallel_{i} L_i)$};
	\node[]		(5) [below of = 4] {};
	\draw[->] (4) -- (5);
	\draw[] (0, -.5) rectangle (4, 0.1);
	\node[]			(3) at (2, -.2) {Decision};
\end{tikzpicture}
\caption{Architecture of a system with centralized diagnosis}
\label{fig_centralized}
\end{figure}

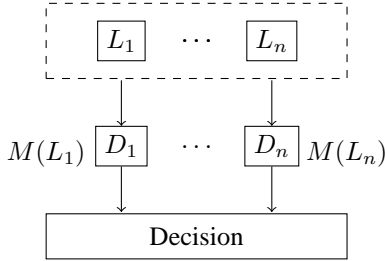
\begin{figure}[t]
\centering
\begin{tikzpicture}
	\node[draw]		(0)	at (1, 3) 		{$L_1$};
	\node[]			(1)	[right of = 0]	{$\ldots$};
	\node[draw]		(2)	[right of = 1]	{$L_n$};
	\draw[dashed] (0, 2.5) rectangle (4, 3.5);

	\node[]			(000)	at (1, 2.6) {};
	\node[draw]		(00) [below of = 000] {$D_1$};
	\node[]			(10) [below of = 00] {$$};
	\draw[->] (000) -- (00);
	\draw[->] (00) -- (10);

	\node[]			(1)	[right of = 0]	{$\ldots$};
	\node[]			(11)[right of = 00]	{$\ldots$};

	\node[]			(222)	at (3, 2.6) {};
	\node[draw]		(22) [below of = 222] {$D_n$};
	\node[]			(20) [below of = 22] {$$};
	\draw[->] (222) -- (22);
	\draw[->] (22) -- (20);

	\node[]	at (0, 1.5) {$M(L_1)$};
	\node[]	at (4, 1.5) {$M(L_n)$};

	\draw[] (0, 0.7) rectangle (4, 0.1);
	\node[]			(3) at (2, .4) {Decision};
\end{tikzpicture}
\caption{Architecture of a system with decentralized diagnosis}
\label{fig_decentralized}
\end{figure}

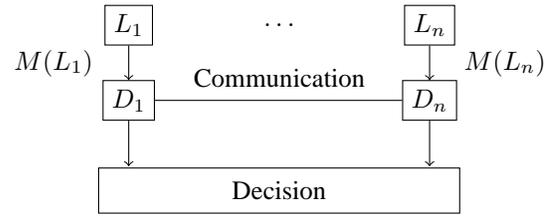
\begin{figure}[t]
\centering
\begin{tikzpicture}
	\node[draw]		(0)	at (1, 2) 		{$L_1$};
	\node[draw]		(00) [below of = 0] {$D_1$};
% 	\node[draw]		(p0) [below of = 00] {Decision$_1$};
	\node[]			(10) [below of = 00] {};
	\draw[->] (0) -- (00);
	\draw[->] (00) -- (10);

	\node[]			(1)	at (3, 2)	{$\ldots$};
	\node[] at (3, 1.3) {Communication};

	\node[draw]		(2)	at (5, 2) 		{$L_n$};
	\node[draw]		(22) [below of = 2] {$D_n$};

	\draw[] (.6, 0.1) rectangle (5.4, -0.5);
	\node[]			(3) at (3, -0.2) {Decision};

% 	\node[draw]		(p2) [below of = 22] {Decision$_n$};
	\node[]			(20) [below of = 22] {};
	\draw[->] (2) -- (22);
	\draw[->] (22) -- (20);
	\draw[] (00) -- (22);

	\node[]	at (0, 1.5) {$M(L_1)$};
	\node[]	at (6, 1.5) {$M(L_n)$};
\end{tikzpicture}
\caption{Architecture of a system with distributed diagnosis}
\label{fig_distributed}
\end{figure}

%%%%%%%%%%%%%%%%%%%%%%%%%%%%%%%%%%%%%%%%%%%%%%%%%%%%%%%%%%%%%%%%%%%%%%%%%%%%%%%
\section{Preliminaries}
\label{sec:Preliminaries}

\subsection{Notation}
The notation used in this document is the one in
\cite{cassandras_introduction_2010}.
Let $\Sigma$ be a finite set of events. A sequence of events is a string.
$\Sigma^*$ denotes a set of all finite strings over $\Sigma$.
$L\subseteq\Sigma^*$ is a language over $\Sigma$. Given strings $s$ and $t$,
$st$ is their concatenation. Given strings $s$ and $w$, $w$ is a prefix of $s$
if exists $t$ such that $wt = s$. Prefix closure of $L$, denoted by
$\overline{L}$ is a set of all prefixes of all the strings in $L$.
If $\overline{L} = L$ then $L$ is prefix-closed. The post language of $L$ after
a string $s$ is denoted as $L/s$, i.e. $L/s := \{t\mid st \in L\}$. We
write $\sigma \in s$ if the event $\sigma \in \Sigma$ appears in the string $s
\in \Sigma^*$. If $\{s\}$ is a singleton, we write $s$ for operations on
languages.

An automaton $G$ is a tuple $$G := \left< X,\Sigma,\delta,x_0, X_m \right>,$$
where $X$ is a set of states, $x_0 \in X$ is an initial state, $X_m \subseteq X$
is the set of marked states, and $\delta: X \times \Sigma \rightarrow X$ is the
transition function.
We say a language $L := \mathcal{L}(G)$ is generated or recognized by the
automaton $G$. In this paper we assume that for each language there is always a
correspondent automaton, and vice versa. The marked language $L_m \subseteq L$
is intended to make a part of the automaton's behaviour distinguishable in a
certain context.

Some events of DES can not be observed. To reflect that the set of events
$\Sigma$ is partitioned into disjointed sets of observable events $\Sigma_o$ and
not observable events $\Sigma_{ou}$, i.e. $\Sigma = \Sigma_o~\dot{\cup}~
\Sigma_{ou}$.
The $M: \Sigma^* \rightarrow \Sigma_o^*$ denotes the natural projection that
erases unobservable events.
% ; $\epsilon$ means the empty string. 
The correspondent inverse projection is $M^{-1}: \Sigma_o^* \rightarrow
2^{\Sigma^*}$.
If a set of events is partitioned into subsets, $\Sigma := \bigcup_i
\Sigma_{i} \mid i \in \mathbb{N}$, the natural projection over the partition
members is denoted as $M_i: \Sigma_i^* \rightarrow \Sigma_{i,o}^*$.

Let $I := \{1,2,\ldots,n\} \subset  \mathbb{N}$ be an index set. A system is
defined by a set of automata $\{G_{i \in I}\}$ and a correspondent set of
languages $\{L_{i \in I}\}$. We use the term \emph{local} in context of the
automata and languages from these sets. The \emph{global} language of the system
is defined by the parallel composition \cite{cassandras_introduction_2010} of
its local languages:
$L := \parallel_{i \in I} L_i$.
The natural projection is commonly defined over Kleene closure on event sets.
We restrict it, for simplicity of notation, to the system's languages as
follows: $P_i(L) := \{s\mid s\in L_{i}\}$, and $P_i^{-1}(L_{i}) := \{s \mid s
\in L\}, ~i \in I$.

\subsection{Architectures for on-line diagnosis}
Architectures for on-line diagnosis can be categorized as follows:
centralized, decentralized and distributed.

\subsubsection{Centralized approach}
This architecture refers to a global model (language).If the system is modular,
then the global language is built by the parallel composition of the local
languages. All the observations are performed at one site. In this architecture
only one diagnoser $D$ \cite{sampath_diagnosability_1995} is constructed. Upon
the current state of the diagnoser a decision on the fault occurrence is made.
The structure is depicted in Figure \ref{fig_centralized}.

\subsubsection{Decentralized approach}
This approach also exploits the entire model built from its modules, but several
local sites perform observations using only local diagnosers.
The diagnosers do not communicate to each other, but they provide necessary
information (via a protocol) to a central decision node. This architecture is
depicted in Figure \ref{fig_decentralized}.

\subsubsection{Distributed approach}
The architecture is depicted in Figure \ref{fig_distributed}.
The distributed approach does not require to built the entire model of the
system. The architecture implies that the system has a set of observation spots,
and each spot observes only one module of the system. A communication among
observation spots is possible in order to make a decision about a fault
occurrence. 

The notion of modular diagnosability meets the same architectural implications,
and we refer to it as to the distributed approach when the amount of information
the observation spots communicate to each other is equal to zero.

%%%%%%%%%%%%%%%%%%%%%%%%%%%%%%%%%%%%%%%%%%%%%%%%%%%%%%%%%%%%%%%%%%%%%%%%%%%%%%%
\section{Diagnosability of a Modular System}
\label{sec:Diagnosability}

%\subsection{Faulty and non-faulty languages}
Diagnosability analysis uses a notion of a faulty language to describe the
faulty behaviour of a discrete-event system. This section discuses design issues
related to representations of the faulty language and focuses on a definition
of modular diagnosability.

The faulty behavior is usually modeled by introducing fault events or by faulty
specifications. We refer to this approaches as to \emph{event-based} and
\emph{specification-based} correspondingly. All the aforementioned works exploit
the event-based approach, whereas the works \cite{zhou_decentralized_2008} and
\cite{sartini_methodology_2010} are examples of the specification-based one.

In the event-based approach fault events are a special type of event such that
$\Sigma_{uo}$ can be disjointed into the sets of faults $\Sigma_f$ and
non-faults $\Sigma_{uo}\backslash \Sigma_f$. A string containing a fault event
is called \emph{faulty string}. A set of faulty strings is called \emph{faulty
language}, i.e. formally 
$$L_f := \{ s \in L \mid \sigma \in s, \sigma \in \Sigma_f\}.$$ 
By definition, the faulty language is not necessarily prefix-closed,
$L_f \subseteq \overline{L_f}$. Thus, in the event-based approach the language
of the system can be partitioned into faulty and non-faulty languages, where the
\emph{non-faulty language} is defined as $L_{nf} := L \backslash L_f$.

In the case of the specification-based approach the faulty specification allows
us to define undesired behavior when the fault events are not necessarily
introduced. In this case this behaviour can be represented by a marked
language $L_f := L_m \subseteq L$. Labeling automata's states for the same
purpose can be considered as an equal technique.

Different types of undesired behaviours (or types of faults) are defined by
partitioning $\Sigma_f$ into subsets (not necessarily disjoint) or by several
faulty specifications for the same language. 

A faulty language defined in the event-based approach can be simply converted
into a faulty specification by marking faulty strings, and erasing fault events.
Then, we can assume that if fault events are defined, then faulty specifications
can also be defined. Consequently, a set of different types of faults requires a
correspondent set of specifications.
Thus, a method suitable for the specification-based approach implies that it can
be adopted for the event-based approach. In this paper, for the sake of
unification, we use specification-based approach. For this reason the
definitions of diagnosability originally developed by their authors for the
event-based approach are slightly modified with no loss of meaning.

For the sake of simplicity, in the following we assume that there is only one
type of fault, and that the language of the system is live.

We define \emph{diagnosability of a fault} as follows:
\begin{definition} 
\label{def:fault_is_diag}
Given a system's language $L$ with a fault defined by the sublanguage $L_f$. The
fault is diagnosable if there is no two strings in the language $L$ with the
same observation such that one string is faulty and of arbitrary cardinality,
and another is non-faulty, i.e. if the following holds:
\end{definition}
\begin{equation}
\begin{array}{l}
	\forall(s \in L_f, t \in L_f/s) 
	\\
	(\exists n \in \mathbb{N})
	(|t| \geq n) 
	\\
	\left[ M(st) \cap M(L_{nf}) = \emptyset \right].
\end{array}
\end{equation}

We define \emph{diagnosability property of a language} as follows:
\begin{definition}
The language is diagnosable if all its faults are diagnosable.
\end{definition}
The two above definitions altogether are similar to the Definition 1 in
\cite{sampath_diagnosability_1995}. We recall the statement in
\cite{contant_diagnosability_2006} proved by Theorem 2, that the global
language of the system is not diagnosable only if exists at least one non-diagnosable
local language. If all the local languages are diagnosable then the global
language is diagnosable. We refer to this property as to a local diagnosability
property:

\begin{definition}[Local diagnosability] Given the set of languages
$\{L_{i \in I}\}$. The global language $L := \parallel L_i$ is
diagnosable locally if each local language $L_i$ is diagnosable, i.e. if
the following holds:
\end{definition}
\begin{equation}
\begin{array}{l}
	\forall(i \in I, s \in L_{i,f}, t \in L_{i,f}/s)
	\\
	(\exists n \in \mathbb{N})
	(|t| \geq n)
	\\
	\left[ M_i(st) \cap M_i(L_{i,nf}) = \emptyset \right].
\end{array}
\end{equation}

The definition of modular diagnosability extends the definition of local
diagnosability as it takes into account the case when a faulty string locally
indistinguishable in one module becomes distinguishable due to the
composition with another module:

\begin{definition}[Modular diagnosability] Given the set of local languages
$\{L_{i \in I}\}$ and its correspondent sets $\{L_{i,f}\}$ and
$\{L_{i,nf}\}$. The global language $L := \parallel L_i$ is \emph{modularly
diagnosable} with respect to
$M_i: \Sigma^* \rightarrow \Sigma_{i,o}^*$ 
if the following holds:
\end{definition}
\begin{equation}
\begin{array}{l}
	\forall(i \in I, s \in L_{i,f}, t \in L_{i,f}/s)
	\\
	(\exists n \in \mathbb{N})
	(|t| \geq n)
	\\
	\left[ M_i(P_i^{-1}(st)) \cap M_i(P_i^{-1}(L_{i,nf})) = \emptyset \right].
\end{array}
\end{equation}

It was proved in \cite{contant_diagnosability_2006} by Theorem 2, Part 2 
that the local diagnosability implies the modular diagnosability\footnote{In
\cite{zhou_decentralized_2008} the authors show that the local diagnosability
and modular diagnosability are not comparable but they have a different setup
for the problem.}, i.e.
\begin{equation}
\begin{array}{l}
	\forall(i \in I, s \in L_{i,f}, t \in L_{i,f}/s)
	\\
	(\exists n \in \mathbb{N})
	(|t| \geq n)
	\\
	\left[
	\left( M_i(st) \cap M_i(L_{i,nf}) = \emptyset \right)
	\Rightarrow \right.
	\\ 
	\left.
	\left( M_i(P_i^{-1}(st)) \cap M_i(P_i^{-1}(L_{i,nf})) = \emptyset \right)
	\right].
\end{array}
\end{equation}

Recall the Definition \ref{def:fault_is_diag} of the diagnosable fault. If a
module is not diagnosable locally then exist at least two strings in its
language, one is faulty and the other one is not, with the same observation of
arbitrary length, i.e. the strings are \emph{not distinguishable}. The
indistinguishability can disappear if and only if:
$a)$ at least one string is not in its language due to concurrency with other
module, and then the strings would be distinguishable locally - the verification
of the modular diagnosability property is devoted to find if this is the case;
$b)$ indistinguishability is broken globally by interleaving sequences of the
module's events with observable events of other modules. The later case is
expressed in the following conjecture:
\begin{conjecture} Given a system of two modules with languages $L_1$ and $L_2$,
and the global language $L := L_1 \parallel L_2$. Suppose there is only
one faulty string $s \in L_1$ such that it is not distinguishable from at least one
string of $L_1\backslash s$. Thus, $L_1$ is not locally diagnosable. Suppose the
system is not modular diagnosable. Then the global language $L$ is diagnosable
only if all the strings $t \in P_1^{-1}(s)$ change their observation due to 
the composition with the language $L_2$.
\end{conjecture}

The above conjecture gives the insight into the underlining idea of our
approach. If we find a module which makes the faulty string distinguishable
then the composition of that module with a faulty one would result in a new
module satisfying the property of local diagnosability, thus improving
the modular diagnosability property of the system. In the following section we
provide a formal description of the problem.

% In the above the infinite length of the string $s$ is equal to the statement
% that the string $s$ forms a cycle. This definition is more weak then one
% presented in \cite{sampath_diagnosability_1995} because it does not assume non
% existence of unobservable cycles.

\section{Virtual Modules and structural Analysis}
\label{sec:Proposal}

Our goal is to have the system modularly diagnosable. If the initial modularity
does not satisfy the property of modular diagnosability then we assume that the
set of modules can be partitioned such that all the modules in each
element of the partition can be considered as a \emph{virtual module}, and the
system with the new modularity satisfies the property of modular diagnosability.

\begin{definition}[Diagnosability of virtual modules] 
Let $I := \{1,2,\ldots,n\} \subset  \mathbb{N}$ be an index set, and $J$ be a
partition of $I$. Given the set of local languages $\{L_{i \in I}\}$ and its
corresponded subsets $\{L_{i,f}\}$ and $\{L_{i,nf}\}$. The global language $L :=
\parallel L_i$ is modularly diagnosable with respect to
$M_j: \Sigma^* \rightarrow \Sigma_{j,o}^* 
\mid j \in J, ~\Sigma_{j,o} :=\bigcup_{i \in j} \Sigma_{i,o}$ 
if the following holds:
\end{definition}
\begin{equation}
\begin{array}{l}
	\forall(i \in I, s \in L_{i,f}, t \in L_{i,f}/s)
	\\
	(\exists n \in \mathbb{N})
	(|t| \geq n)
	\\
	\left[ M_j(P_i^{-1}(st)) \cap M_j(P_i^{-1}(L_{i,nf})) = \emptyset \right].
\end{array}
\end{equation}

If $\forall j \in J, L_{j} := \parallel L_{i \in j}$ then, by definition, $L :=
\parallel L_{i \in I} := \parallel L_{j \in J}$, and all the statements related
to the modular diagnosability  property can be applied for the diagnosability by
virtual modules.

The diagnosability problem with virtual modules can be solved in two ways.
In the first, in order to find a partition of system's modules satisfying the
modular diagnosability property one may take a faulty module, enumerate all possible
sets of other modules, compose all the languages from each set with the faulty
language, and check the resulting language for modular diagnosability. Since, in
general, there may be many partitions such that the system is diagnosable with
virtual modules, only one partition should be chosen taking some
heuristic guiding criteria. The entire process is computationally expensive,
since the number of possible partitions $J$ is double exponential with respect
to the cardinality of $I$.
However, not each module can change diagnosability. Consequently, we propose the
second approach that can significantly decrease the complexity by selecting only
the modules which can probably change diagnosability and, thus, check only the
partitions made of such modules. For this purpose a procedure to check if an
arbitrary module potentially can change diagnosability is required. Then, we can
have a heuristic procedure to choose which module to pick to verify
diagnosability.

In the sequel, for the sake of simplicity, we suppose that the system consists
only of two modules with the correspondent languages $L_1$ and $L_2$. The
language of the system is $L := L_1 \parallel L_2$. Suppose that only one module
has a faulty behaviour: $L_1 := L_{1,f} ~\dot{\cup}~ L_{1,nf}$.
Suppose that $L_1$ is not diagnosable locally, but $L$ is diagnosable.

Firstly, we define the notion of observation changing of a string in a global
language.
\begin{definition}Given two languages $L_1$ and $L_2$. A string $s \in
L_1$ \emph{changes its observation} $M_1(s)$ in the language $L$ if
there is no the same observation in $P_1^{-1}(s)$, i.e.
if the following holds:
\end{definition}
\begin{equation}
\label{def:obs}
\begin{array}{l}
	M(L) \cap M_1(s) = \emptyset.
\end{array}
\end{equation}

\begin{lemma}
\label{lem_changed_observation}
Given two languages $L_1$ and $L_2$, and a string $s \in L_1$.
Assume that $s \in P_1(L)$. The string $s$ changes its observation in the
language $L$ if and only if:
\end{lemma}
\begin{subequations}\label{lem:obs}
\begin{align}
	(\exists \sigma \in s \mid \sigma \in \Sigma_1 \cap \Sigma_2) \land
	\label{lem:obs1}
	\\
	(\forall t\sigma \in L_2)
	\left[M_2(t) \neq \emptyset \right],
	\label{lem:obs2}
	\\
	\textrm{where } M_2: \Sigma^* \rightarrow (\Sigma_{2,o} \backslash
	\Sigma_1)^*. 
	\label{lem:obs3}
\end{align}
\end{subequations}

\begin{proof}
In order to prove sufficiency of (\ref{lem:obs}) we use its converse relation
and prove by contradiction that the change of observation (\ref{def:obs}) is
necessary.
Assume $\exists w \in L$ and $\exists s \in L_1$ such that
$M(w)= M_1(s)$ and, therefor, (\ref{def:obs}) is false. Let $\exists
\sigma \in \Sigma_1 \cap \Sigma_2$ such that $\sigma \in w$ and also
(\ref{lem:obs1}) holds. Then may $\exists u \sigma \in \overline{w}$ such that
$M_2(u) = \emptyset$, and then $M_2(P_2(u)) =
\emptyset$ which contradicts (\ref{lem:obs2}). Now, let (\ref{lem:obs2})
be true for all $t\sigma \in P_2(w)$. Then the assumption $M(w)=
M_1(s)$ holds only if $\sigma \not \in s$, which contradicts (\ref{lem:obs1}). 

We prove necessity of (\ref{lem:obs}) by contradiction.
Let (\ref{lem:obs1}) holds, and $\exists t\sigma \in L_2$ such that
$M_2(t) = \emptyset$. Then may $\exists t'\sigma \in 
L \subseteq P_2^{-1}(t\sigma)$ such that $M_2(t') =
\emptyset$ and $M_1(t') = M_1(s) \neq \emptyset$, which contradicts
(\ref{def:obs}).
Now, let (\ref{lem:obs2}) holds and $\not \exists \sigma \in s' \in P_1^{-1}(s)
\mid \sigma \in \Sigma_1 \cap \Sigma_2$. Then may $\exists w \in L_2$ and,
hence, $w' \in L \subseteq P_2^{-1}(w)$ such that $M(w')=M(s)$,
which contradicts (\ref{def:obs}).
\end{proof}

Informally, the above lemma says that the string of the local language $L_1$
changes its observation in the global language $L$ if and only if the string has
an event in common with the language $L_2$, and all the strings of $L_2$ which
have this common event have observable events in the prefixes, and some of the
observable events in the prefixes are not common with $L_1$.

We call the subset of stings $\{t \in L_2\}$ satisfying condition
(\ref{lem:obs2}) as the \emph{adjacent observable support} for the given string
$s \in L_1$.

\begin{definition} Given two languages $L_1$ and $L_2$. We say that a string $s
\in L_{1}$ is distinguished from all the other local strings $L_1\backslash s$
in the language $L$ if the following holds:
\end{definition}
\begin{equation}
\label{def:dist}
\begin{array}{l}
% 		M(s \parallel L_2) \cap M((L_1\backslash s) \parallel L_2)
% 		= \emptyset.
  (\forall w \in L_1\backslash s)\\
  	\left[ MP_1^{-1}(w \parallel L_2) \cap MP_1^{-1}(s\parallel L_2) = \emptyset
  	\right].
\end{array}
\end{equation}

\begin{lemma}
\label{lem:distinguished}
Given two languages $L_1$ and $L_2$. Assume that $L_1 = P_1(L)$. The string $s
\in L_1$ is distinguished from $L_1\backslash s$ in the language $L$ if $s$ has
an adjacent observable support $L_{2,s} \subseteq L_2$ which
satisfies the following condition:
\end{lemma}
\begin{subequations}
\begin{align}
	(\forall t \in L_{2,s})
	\left[ \exists \sigma \in t \mid \sigma \in \Sigma_1 \cap \Sigma_2)\right]
	\land
	\label{lem:dist1}
	\\
	(\forall w \in L_1\backslash s)\left[\sigma \not \in w \right]
	\label{lem:dist2} \land
	\\
	(\forall t'\sigma \in \overline{t})
	[M_2(t / t'\sigma) \neq \emptyset]
	\label{lem:dist3},
\end{align}
\end{subequations}

where $M_2$ is defined as in (\ref{lem:obs3}). 

\begin{proof}
Assume (\ref{def:dist}) is false, i.e. $\exists w' \in P_1^{-1}(w)$ and $\exists
s' \in P_1^{-1}(s)$ such that $M(w') = M(s')$. 

Assume (\ref{lem:dist1}) and (\ref{lem:dist2}) hold. Then may $\exists t' \in
\overline{s'} \mid t' \in P_2^{-1}(L_{2,s})$ such that $M(t') = M(w')$, and $t
\in P_2(t')$ such that $M_2(t) = M_2(w)$. And may $\exists t'' \in \overline{t}$
such that $M_2(t'') = M_2(w)$. Since $\sigma \in t$ and $\sigma \not \in w$,
then $M(t \backslash t''\sigma) = \emptyset$ for any $t''\sigma \in
\overline{t}$, which contradicts (\ref{lem:dist3}).

Assume (\ref{lem:dist1}) and (\ref{lem:dist3}) hold. Let $M_1(L_1) = \emptyset$
and $M_2(t\backslash t'\sigma) = M_2(s) = M_2(s)$. Then $\forall s' \in
M(P_1^{-1}(s))$ there exists $\sigma \in s'$, which contradicts
(\ref{lem:dist2}).

Assume (\ref{lem:dist2}) and (\ref{lem:dist3}) hold. If (\ref{def:dist}) is
false, then (\ref{lem:dist1}) is false. However, (\ref{lem:dist3}) is sufficient
for (\ref{lem:dist1}), which contradicts the former statement.
\end{proof}

Informally, the above lemma says that a string $s$ becomes distinguishable from
the other strings $L_1\backslash s$ in the global language, when the occurrence
of events from the observable support happens only in $P^{-1}(s)$ due to
common events. Thus, whenever we observe events of the observable support of
$L_2$, we are sure the string $s$ in $L_1$ is being executed.

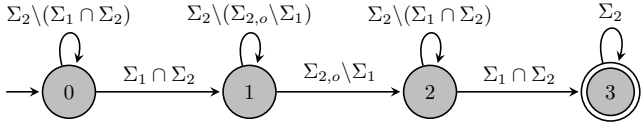
\begin{figure}[t]
\centering
\begin{tikzpicture}
[->,>=stealth, node distance=30mm, auto, initial text=, shorten >=1pt,
semithick, 
every state/.style={fill=lightgray},
every node/.style={scale=0.8}, 
accepting/.style={double distance=1.5pt, outer sep=0.75pt+\pgflinewidth}
]
  \node[initial,state] (0)              {$0$};
  \node[state]         (1) [right of=0] {$1$};
  \node[state]         (2) [right of=1] {$2$};
  \node[state, accepting]         (3) [right of=2] {$3$};

  \path (0) edge [loop above] node {$\Sigma_2 \backslash (\Sigma_1 \cap \Sigma_2)$} (0)
  		(0) edge              node {$\Sigma_1 \cap \Sigma_2$} (1)
  		(1) edge [loop above] node {$\Sigma_2 \backslash (\Sigma_{2,o} \backslash \Sigma_1)$} (1)
        (1) edge              node {$\Sigma_{2,o} \backslash \Sigma_1$} (2)
  		(2) edge [loop above] node {$\Sigma_2 \backslash (\Sigma_1 \cap \Sigma_2)$} (2)
        (2) edge              node {$\Sigma_1 \cap \Sigma_2$} (3)
		(3) edge [loop above] node {$\Sigma_2$} (3)
%         (2) edge [bend left]  node {$\Sigma_1 \cap \Sigma_2$} (1)
        ;
\end{tikzpicture}
\caption{Automaton for marking the language $L_2$}
\label{fig:marking_L2}
\end{figure}

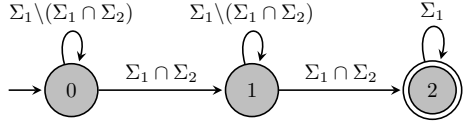
\begin{figure}[t]
\centering
\begin{tikzpicture}
[->,>=stealth, node distance=30mm, auto, initial text=, shorten >=1pt,
semithick, 
every state/.style={fill=lightgray},
every node/.style={scale=0.8}, 
accepting/.style={double distance=1.5pt, outer sep=0.75pt+\pgflinewidth}
]
  \node[initial,state] (0)              {$0$};
  \node[state]         (1) [right of=0] {$1$};
  \node[state, accepting]         (2) [right of=1] {$2$};

  \path 
	(0) edge [loop above] node {$\Sigma_1 \backslash (\Sigma_1 \cap \Sigma_2)$} (0)
	(0) edge              node {$\Sigma_1 \cap \Sigma_2$} (1)
	(1) edge [loop above] node {$\Sigma_1 \backslash (\Sigma_1 \cap \Sigma_2)$} (1)
	(1) edge              node {$\Sigma_1 \cap \Sigma_2$} (2)
	(2) edge [loop above] node {$\Sigma_1$} (2)
% 	(1) edge [bend left]  node {$\Sigma_1 \cap \Sigma_2$} (0)
    ;
\end{tikzpicture}
\caption{Automaton for marking the language $L_1$}
\label{fig:marking_L1}
\end{figure}

As it was discussed in the Section \ref{sec:Diagnosability},
indistinguishability can be changed either by blocking the string in the local
language due to concurrency, or by interleaving with observable events from
other languages. Under assumption that all the strings are not affected by
concurrency, i.e. $L_1 = P_1(L)$ we can deduce, that the conditions of Lemma
\ref{lem:distinguished} are also necessary for changing distinguishability. 

The Figure \ref{fig:marking_L2} depicts an automaton which
accepts the sublanguage of $L_2$ satisfying conditions (\ref{lem:obs1}),
(\ref{lem:dist1}) and (\ref{lem:dist3}) of the above lemmas.
The Figure \ref{fig:marking_L1} depicts an automaton which
marks a sublanguage of $L_1$ satisfying conditions (\ref{lem:obs1}) and
(\ref{lem:dist1}). 

A procedure verifying if a string $s \in L_1$ is distinguishable in the global
language $L$ consists of two steps. First, the string $s$ should be marked by
the automaton depicted in the Figure \ref{fig:marking_L1}. Then the set of
common events $\Sigma_1 \cap \Sigma_2$ is reduced to the set of events causing
transitions in the automaton. Second, all the continuations of the strings of
the language $L_2$ which have these common events should be accepted by the
automaton depicted in the Figure
\ref{fig:marking_L2}.

Now we are ready to apply Lemma \ref{lem:distinguished} with respect to
diagnosability property, but make some notes before. Intuitively, one would
apply the conditions of the lemma for faulty and non-faulty languages. Recall,
that faulty and non-faulty languages are disjoint, but they may have common
prefixes. This common sublanguage is defined as $\overline{L_{f}} \cap (L
\backslash \overline{L_{f}})$. Changing observability of this sublanguage has no
effect for diagnosability, and we can exclude it from a verification procedure.
Thus, the non-faulty sublanguage disjoint to all the prefixes of the faulty
language is defined as $L \backslash \overline{L_f}$, and the set of all
prefixes of the faulty language disjoint to the above non-faulty sublanguage and
to the common prefixes is defined as $\overline{L_f} \backslash (\overline{L_f}
\cap (L \backslash \overline{L_f}))$.

\begin{lemma}
\label{lem:diagnosable}
Given $L_1, L_2$, $L_{1,f} \subseteq L_1$ and $L_{1,nf} \subseteq L_1$. A
language $L := L_1 \parallel L_2$ is diagnosable if the sublanguages
$\overline{L_{1,f}} \backslash (\overline{L_{1,f}} \cap (L_i \backslash
\overline{L_{1,f}}))$ and $L_{1} \backslash \overline{L_{1,f}}$ have
distinguished observable supports in $L_2$.
% , which satisfy conditions \ref{lem:dist1}-\ref{lem:dist3}.
\end{lemma}
The proof can be deduced from the Lemma \ref{lem:distinguished}. 

The automata depicted in Figures \ref{fig:marking_L2} and
\ref{fig:marking_L1} can not be simply used in a procedure verifying
diagnosability, since we should avoid the verification of $\overline{L_{1,f}}
\backslash L_{1,f}$ and $L_{1,nf} \backslash \overline{L_{1,f}}$. We leave
development of such procedure for future work. However, the automata can be
used to demonstrate the approach in a trivial case, as it is shown in the next
section.

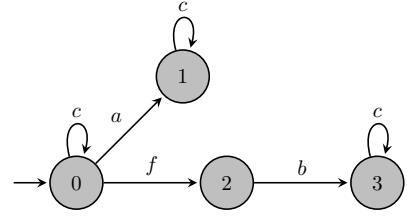
\begin{figure}[t]
\centering
\begin{tikzpicture}
[->,>=stealth, node distance=25mm, auto, initial text=, shorten >=1pt,
semithick, 
every state/.style={fill=lightgray},
every node/.style={scale=0.8}, 
accepting/.style={double distance=1.5pt, outer sep=0.75pt+\pgflinewidth}
]
  \node[initial,state]		(0)              		{$0$};
  \node[state ]   (1) [above right of=0]	{$1$};
  \node[state]         		(2) [right of=0]		{$2$};
  \node[state]	(3) [right of=2]		{$3$};

  \path
 	(0) edge [loop above] node {$c$} (0)
 	(0) edge              node {$a$} (1)
 	(1) edge [loop above] node {$c$} (1)
 	(0) edge              node {$f$} (2)
 	(2) edge              node {$b$} (3)
 	(3) edge [loop above] node {$c$} (3)
    ;
\end{tikzpicture}
\caption{Automaton $G_1$}
\label{fig:G1}
\end{figure}

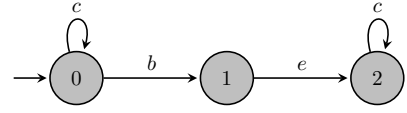
\begin{figure}[t]
\centering
\begin{tikzpicture}
[->,>=stealth, node distance=25mm, auto, initial text=, shorten >=1pt,
semithick, 
every state/.style={fill=lightgray},
every node/.style={scale=0.8}, 
accepting/.style={double distance=1.5pt, outer sep=0.75pt+\pgflinewidth}
]
  \node[initial,state]		(0)              		{$0$};
  \node[state]         		(1) [right of=0]		{$1$};
  \node[state]	(2) [right of=1]		{$2$};

  \path
%  	(0) edge              node {$a$} (1)
 	(0) edge [loop above] node {$c$} (0)
 	(0) edge              node {$b$} (1)
 	(1) edge              node {$e$} (2)
 	(2) edge [loop above] node {$c$} (2)
    ;
\end{tikzpicture}
\caption{Automaton $G_2$}
\label{fig:G2}
\end{figure}

%%%%%%%%%%%%%%%%%%%%%%%%%%%%%%%%%%%%%%%%%%%%%%%%%%%%%%%%%%%%%%%%%%%%%%%%%%%%%%%
\section{Example}
\label{sec:Example}

\begin{figure}[t]
\centering
\begin{tikzpicture}
[->,>=stealth, node distance=22mm, auto, initial text=, shorten >=1pt,
semithick, 
every node/.style={scale=0.8}, 
accepting/.style={double distance=1.5pt, outer sep=0.75pt+\pgflinewidth}
]
  \node[draw, initial](0N0N)              	  {$0N;0N$};
  \node[draw]         (2F0N) [below of =0N0N] {$2F;0N$};
  \node[draw]         (1N0N) [right of =2F0N] {$1N;0N$};
  \node[draw]         (3F0N) [below of =2F0N] {$3F;0N$};
  \node[draw]         (2F2F) [left of =3F0N]  {$2F;2F$};
  \node[draw]         (1N2F) [right of =3F0N] {$1N;2F$};
  \node[draw]         (1N1N) [right of =1N2F] {$1N;1N$};
  \node[draw]         (1N3F) [below of =1N2F] {$1N;3F$};
  \node[draw]         (3F2F) [below of =3F0N] {$3F;2F$};
  \node[draw]         (3F3F) [below of =3F2F] {$3F;3F$};

  \path 
	(0N0N) edge	node {$f$} (2F0N)
	(0N0N) edge	node {$a$} (1N0N)
	(2F0N) edge	node {$f$} (2F2F)
	(2F0N) edge	node {$f$} (2F2F)
	(2F0N) edge	node {$b$} (3F0N)
	(2F0N) edge	node {$a$} (1N2F)
	(1N0N) edge	node {$f$} (1N2F)
	(1N0N) edge	node {$a$} (1N1N)
	(1N1N) edge [loop below] node {$c$} (1N1N)
	(2F2F) edge	node {$b$} (3F2F)
	(3F0N) edge	node {$f$} (3F2F)
	(3F0N) edge	node {$a$} (1N3F)
	(1N2F) edge	node {$b$} (1N3F)
	(3F2F) edge	node {$b$} (3F3F)
	(3F3F) edge [loop below] node {$c$} (3F3F)
	(1N3F) edge [loop below] node {$c$} (1N3F)
    ;
\end{tikzpicture}
\caption{Verifier of $G_1$}
\label{fig:verifier_G1}
\end{figure}
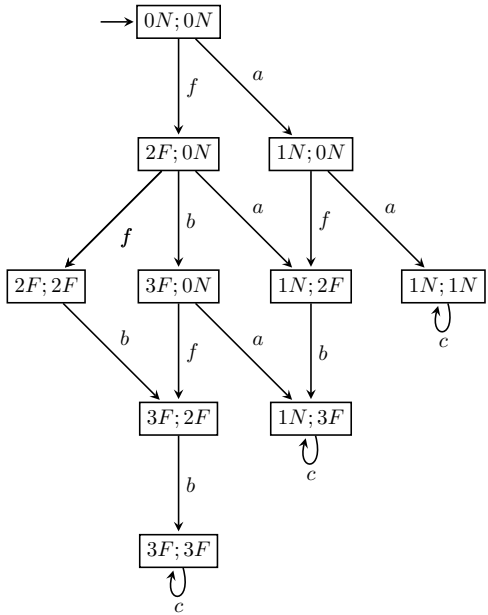

Consider the system of two automata $G_1$ and $G_2$ depicted in Figure
\ref{fig:G1} and Figure \ref{fig:G2}. The set of events for the system is
$\Sigma = \{a, b, c, e, f\}$. Suppose the observable events are $\Sigma_o = \{c,
e\}$, and the set of fault events is $\{f\}$. Thus, only the language $L_1$ has a
fault, and $L_2$ has not. We use the verifier \cite{yoo_polynomial-time_2002} to
check if a language is diagnosable. The verifier for the language $L_1$ is
depicted in the Figure \ref{fig:verifier_G1}. One can check that it has an
indeterminate cycle.
The strings $fbc^*$ and $ac^*$ are not distinguishable in the local language
$L_1$. Hence, $L_1$ is not locally diagnosable.

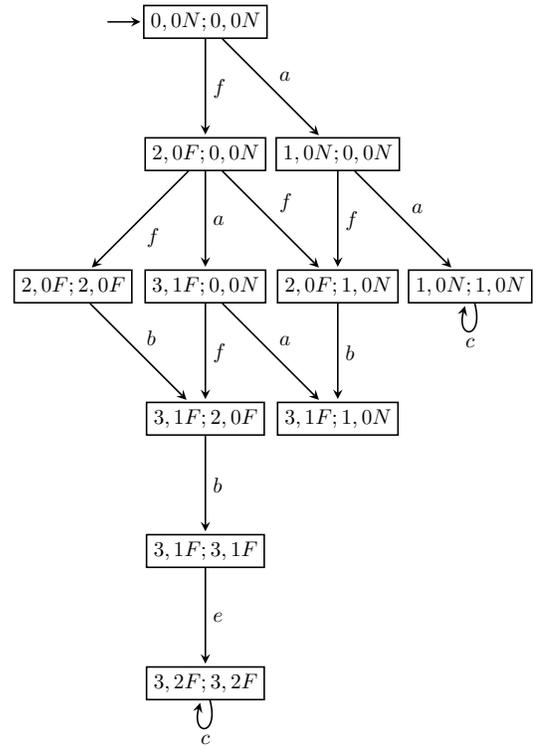
\begin{figure}[t]
\centering
\begin{tikzpicture}
[->,>=stealth, node distance=22mm, auto, initial text=, shorten >=1pt,
semithick, 
every node/.style={scale=0.8}, 
accepting/.style={double distance=1.5pt, outer sep=0.75pt+\pgflinewidth}
]
  \node[draw, initial](00N00N)              	  {$0,0N;0,0N$};
  \node[draw]         (20F00N) [below of =00N00N] {$2,0F;0,0N$};
  \node[draw]         (10N00N) [right of =20F00N] {$1,0N;0,0N$};
  \node[draw]         (31F00N) [below of =20F00N] {$3,1F;0,0N$};
  \node[draw]         (20F20F) [left of =31F00N]  {$2,0F;2,0F$};
  \node[draw]         (20F10N) [right of =31F00N] {$2,0F;1,0N$};
  \node[draw]         (10N10N) [right of =20F10N] {$1,0N;1,0N$};
  \node[draw]         (31F20F) [below of =31F00N] {$3,1F;2,0F$};
  \node[draw]         (31F10N) [below of =20F10N] {$3,1F;1,0N$};
  \node[draw]         (31F31F) [below of =31F20F] {$3,1F;3,1F$};
  \node[draw]         (32F32F) [below of =31F31F] {$3,2F;3,2F$};

  \path
	(00N00N) edge	node {$f$} (20F00N)
	(00N00N) edge	node {$a$} (10N00N)
	(20F00N) edge	node {$f$} (20F20F)
	(20F00N) edge	node {$f$} (20F10N)
	(20F00N) edge	node {$a$} (31F00N)
	(10N00N) edge	node {$f$} (20F10N)
	(10N00N) edge	node {$a$} (10N10N)
	(10N10N) edge [loop below] node {$c$} (10N10N)
	(20F20F) edge	node {$b$} (31F20F)
	(31F00N) edge	node {$f$} (31F20F)
	(31F00N) edge	node {$a$} (31F10N)
	(20F10N) edge	node {$b$} (31F10N)
	(31F20F) edge	node {$b$} (31F31F)
	(31F31F) edge	node {$e$} (32F32F)
	(32F32F) edge [loop below] node {$c$} (32F32F)
    ;
\end{tikzpicture}
\caption{Verifier of $G_1 \parallel G_2$}
\label{fig:verifier_G1G2}
\end{figure}

We now use the verification procedure described in this paper to check if the
language $L_2$ can changes observation of either strings $fbc^*$ or
$ac^*$ in the language $L_1 \parallel L_2$ such that the strings become
distinguishable. The set of events common for $L_1$ and $L_2$ is $\{b, c\}$. It
can be verified that only the strings $fbc^*$ are marked by the automaton
depicted in the Figure \ref{fig:marking_L1}. Next, it can be verified that all
the strings of $L_2$ which have events common with the strings $fbc^*$ are
accepted by the automaton depicted in the Figure \ref{fig:marking_L2}. Thus, we
conclude that $L_2$ changes observation of the strings $fbc^*$ in the virtual
module $G$ built of modules $G_1$ and $G_2$, such that $G$ becomes diagnosable.
Indeed, if we make a parallel composition of the modules and build a verifier
for the result as it is depicted in the Figure \ref{fig:verifier_G1G2}, it can
be checked that the verifier has no indeterminate cycles.

% \begin{figure}[t]
% \centering
%  	\includegraphics[height=80mm]{verifier_G1.jpg}
% \caption{Verifier of $G_1$}
% \label{fig:verifier_G1}
% \end{figure}

% \begin{figure}[t]
% \centering
% \includegraphics[height=100mm]{verifier_G1G2.jpg}
% \caption{Verifier of $G_1 \parallel G_2$}
% \label{fig:verifier_G1G2}
% \end{figure}

% \addtolength{\textheight}{-20.0cm}
%%%%%%%%%%%%%%%%%%%%%%%%%%%%%%%%%%%%%%%%%%%%%%%%%%%%%%%%%%%%%%%%%%%%%%%%%%%%%%%
\section{Conclusion}
\label{sec:Conclusion}

In this paper we introduced a notion of virtual modules for DES, and proposed a
new definition of modular diagnosability by virtual modules. The approach
suggests to combine the existing modules of the system into virtual modules in
a way that the system with the new modularity is modular diagnosable.

We introduced a structural analysis of the system's modules, which allows
to verify if a module may change its observation by composition with others,
and if a module can change the observation of other modules. We defined
correspondent sufficient and necessary conditions for the modules' languages. If
the languages satisfy those conditions then one can state that the system can
be made modular diagnosable by creating virtual modules.
The suggested verification procedure has linear complexity with respect to the
number of states of a module.

We are actually working on the problem of defining criteria of how to select the
best candidates for creating the virtual modules, what the optimal partition of
the set of the modules can be, and generalization of the problem.

% This command serves to balance the column lengths on the last page of the
% document manually. It shortens the textheight of the last page by a suitable
% amount. This command does not take effect until the next page so it should come
% on the page before the last. Make sure that you do not shorten the textheight
% too much.

\bibliographystyle{IEEEtran}
\bibliography{References}

\begin{thebibliography}{10}
\providecommand{\url}[1]{#1}
\csname url@samestyle\endcsname
\providecommand{\newblock}{\relax}
\providecommand{\bibinfo}[2]{#2}
\providecommand{\BIBentrySTDinterwordspacing}{\spaceskip=0pt\relax}
\providecommand{\BIBentryALTinterwordstretchfactor}{4}
\providecommand{\BIBentryALTinterwordspacing}{\spaceskip=\fontdimen2\font plus
\BIBentryALTinterwordstretchfactor\fontdimen3\font minus
  \fontdimen4\font\relax}
\providecommand{\BIBforeignlanguage}[2]{{%
\expandafter\ifx\csname l@#1\endcsname\relax
\typeout{** WARNING: IEEEtran.bst: No hyphenation pattern has been}%
\typeout{** loaded for the language `#1'. Using the pattern for}%
\typeout{** the default language instead.}%
\else
\language=\csname l@#1\endcsname
\fi
#2}}
\providecommand{\BIBdecl}{\relax}
\BIBdecl

\bibitem{su_global_2005}
R.~Su and W.~Wonham, ``Global and local consistencies in distributed fault
  diagnosis for discrete-event systems,'' \emph{{IEEE} Transactions on
  Automatic Control}, vol.~50, no.~12, pp. 1923--1935, 2005.

\bibitem{sampath_diagnosability_1995}
M.~Sampath, R.~Sengupta, S.~Lafortune, K.~Sinnamohideen, and D.~Teneketzis,
  ``Diagnosability of discrete-event systems,'' \emph{{IEEE} Transactions on
  Automatic Control}, vol.~40, no.~9, pp. 1555--1575, Sep. 1995.

\bibitem{jiang_polynomial_2001}
S.~Jiang, Z.~Huang, V.~Chandra, and R.~Kumar, ``A polynomial algorithm for
  testing diagnosability of discrete-event systems,'' \emph{{IEEE} Transactions
  on Automatic Control}, vol.~46, no.~8, pp. 1318 --1321, Aug. 2001.

\bibitem{yoo_polynomial-time_2002}
T.-S. Yoo and S.~Lafortune, ``Polynomial-time verification of diagnosability of
  partially observed discrete-event systems,'' \emph{{IEEE} Transactions on
  Automatic Control}, vol.~47, no.~9, pp. 1491 -- 1495, Sep. 2002.

\bibitem{debouk_coordinated_1998}
R.~Debouk, S.~Lafortune, and D.~Teneketzis, ``Coordinated decentralized
  protocols for failure diagnosis of discrete event systems,'' in \emph{1998
  {IEEE} International Conference on Systems, Man, and Cybernetics, 1998},
  vol.~3, 1998, pp. 3010--3011 vol.3.

\bibitem{pencole_formal_2005}
Y.~Pencolé and M.-O. Cordier, ``A formal framework for the decentralised
  diagnosis of large scale discrete event systems and its application to
  telecommunication networks,'' \emph{Artificial Intelligence}, vol. 164, no.
  1–2, pp. 121--170, May 2005.

\bibitem{qiu_decentralized_2006}
W.~Qiu and R.~Kumar, ``Decentralized failure diagnosis of discrete event
  systems,'' \emph{{IEEE} Transactions on Systems, Man and Cybernetics, Part A:
  Systems and Humans}, vol.~36, no.~2, pp. 384 -- 395, Mar. 2006.

\bibitem{wang_diagnosis_2007}
Y.~Wang, T.-S. Yoo, and S.~Lafortune, ``Diagnosis of discrete event systems
  using decentralized architectures,'' \emph{Discrete Event Dynamic Systems},
  vol.~17, no.~2, pp. 233--263, 2007.

\bibitem{su_distributed_2002}
R.~Su, W.~Wonham, J.~Kurien, and X.~Koutsoukos, ``Distributed diagnosis for
  qualitative systems,'' in \emph{Sixth International Workshop on Discrete
  Event Systems, 2002. Proceedings}, 2002, pp. 169--174.

\bibitem{contant_diagnosability_2006}
O.~Contant, S.~Lafortune, and D.~Teneketzis, ``Diagnosability of discrete event
  systems with modular structure,'' \emph{Discrete Event Dynamic Systems},
  vol.~16, no.~1, pp. 9--37, Jan. 2006.

\bibitem{ye_optimized_2010}
L.~Ye and P.~Dague, ``An optimized algorithm for diagnosability of
  component-based systems,'' in \emph{10th International Conference on Discrete
  Event Systems WODES'10, 2008}, Aug. 2008.

\bibitem{cassandras_introduction_2010}
C.~G. Cassandras and S.~Lafortune, \emph{Introduction to Discrete Event
  Systems}, 2nd~ed.\hskip 1em plus 0.5em minus 0.4em\relax Springer, Oct. 2010.

\bibitem{zhou_decentralized_2008}
C.~Zhou, R.~Kumar, and R.~Sreenivas, ``Decentralized modular diagnosis of
  concurrent discrete event systems,'' in \emph{9th International Workshop on
  Discrete Event Systems, 2008. {WODES} 2008}, May 2008, pp. 388 --393.

\bibitem{sartini_methodology_2010}
M.~Sartini, A.~Paoli, R.~Hill, and S.~Lafortune, ``A methodology for modular
  model-building in discrete automation,'' in \emph{2010 {IEEE} Conference on
  Emerging Technologies and Factory Automation {(ETFA)}}, Sep. 2010, pp. 1 --8.

\end{thebibliography}

\end{document}